\documentclass[]{spie}  %>>> use for US letter paper
\usepackage{amsmath,amsfonts,amssymb}
\usepackage{graphicx}
\usepackage[colorlinks=true, allcolors=blue]{hyperref}

\newcommand{\PSUAA}{Department of Astronomy \& Astrophysics, 525 Davey Laboratory, The Pennsylvania State University, University Park, PA, 16802, USA}
\newcommand{\PSUCEHW}{Center for Exoplanets and Habitable Worlds, 525 Davey Laboratory, The Pennsylvania State University, University Park, PA, 16802, USA}
\newcommand{\PSETI}{Penn State Extraterrestrial Intelligence Center, 525 Davey Laboratory, The Pennsylvania State University, University Park, PA, 16802, USA}
\newcommand{\UA}{Steward Observatory, The University of Arizona, 933 N.\ Cherry Ave, Tucson, AZ 85721, USA}
\newcommand{\NOAO}{NSF's National Optical-Infrared Astronomy Research Laboratory, 950 N.\ Cherry Ave., Tucson, AZ 85719, USA}
\newcommand{\Princeton}{Department of Astrophysical Sciences, Princeton University, 4 Ivy Lane, Princeton, NJ 08540, USA}

\newcommand{\STScI}{Space Telescope Science Institute, 3700 San Martin Dr, Baltimore, MD 21218, USA}
\newcommand{\JHU}{Department of Physics and Astronomy, Johns Hopkins University, 3400 N Charles St, Baltimore, MD 21218, USA}
\newcommand{\GoddardESAL}{Exoplanets and Stellar Astrophysics Laboratory, NASA Goddard Space Flight Center, Greenbelt, MD 20771, USA}
\newcommand{\Macquarie}{Department of Physics and Astronomy, Macquarie University, Balaclava Road, North Ryde, NSW 2109, Australia }
\newcommand{\UCI}{Department of Physics \& Astronomy, The University of California, Irvine, Irvine, CA 92697, USA}
\newcommand{\Carleton}{Carleton College, One North College St., Northfield, MN 55057, USA}
\newcommand{\JPL}{Jet Propulsion Laboratory, California Institute of Technology, 4800 Oak Grove Drive, Pasadena, California 91109}
\newcommand{\TIFR}{Department of Astronomy and Astrophysics, Tata Institute of Fundamental Research, Homi Bhabha Road, Colaba, Mumbai 400005, India}

\title{Real-time exposure control and instrument operation with the NEID spectrograph GUI}

\author[a,b]{Arvind F.\ Gupta}
\author[c]{Chad F.\ Bender}
\author[d]{Joe P.\ Ninan}
\author[e]{Sarah E.\ Logsdon}
\author[a,b]{Shubham Kanodia}

\author[e]{Eli Golub}
\author[e]{Jesus Higuera}
\author[e]{Jessica Klusmeyer}

\author[f]{Samuel Halverson}
\author[a,b]{Suvrath Mahadevan}
\author[g]{Michael McElwain}
\author[h]{Christian Schwab}
\author[i]{Gudmundur Stefansson}
\author[j]{Paul Robertson}
\author[k,l]{Arpita Roy}
\author[m]{Ryan Terrien}
\author[a,b,n]{Jason Wright}

\affil[a]{\PSUAA}
\affil[b]{\PSUCEHW}
\affil[c]{\UA}
\affil[d]{\TIFR}
\affil[e]{\NOAO}

\affil[f]{\JPL}
\affil[g]{\GoddardESAL} 
\affil[h]{\Macquarie}
\affil[i]{\Princeton}
\affil[j]{\UCI}
\affil[k]{\STScI}
\affil[l]{\JHU}
\affil[m]{\Carleton}
\affil[n]{\PSETI}

\authorinfo{Contact: arvind@psu.edu}

\begin{document} 
\maketitle

\begin{abstract}
The NEID spectrograph on the WIYN 3.5-m telescope at Kitt Peak has completed its first full year of science operations and is reliably delivering sub-m/s precision radial velocity measurements. The NEID instrument control system uses the TIMS package (Bender et al. 2016), which is a client-server software system built around the twisted python software stack. During science observations, interaction with the NEID spectrograph is handled through a pair of graphical user interfaces (GUIs), written in PyQT, which wrap the underlying instrument control software and provide straightforward and reliable access to the instrument. Here, we detail the design of these interfaces and present an overview of their use for NEID operations.
Observers can use the NEID GUIs to set the exposure time, signal-to-noise ratio (SNR) threshold, and other relevant parameters for observations, configure the calibration bench and observing mode, track or edit observation metadata, and monitor the current state of the instrument. These GUIs facilitate automatic spectrograph configuration and target ingestion from the nightly observing queue, which improves operational efficiency and consistency across epochs.  By interfacing with the NEID exposure meter, the GUIs also allow observers to monitor the progress of individual exposures and trigger the shutter on user-defined SNR thresholds. In addition, inset plots of the instantaneous and cumulative exposure meter counts as each observation progresses allow for rapid diagnosis of changing observing conditions as well as guiding failure and other emergent issues.
\end{abstract}


\section{INTRODUCTION}
\label{sec:intro}

In recent years, new spectrographs have demonstrated the capacity to deliver velocity measurements with sub-m s$^{-1}$ instrumental precision\cite{Blackman2020}, opening the door for extreme precision radial velocity (EPRV) science. 
Among these is NEID\cite{Schwab2016,Robertson2019}, a high spectral resolution (R $\sim120,000$) optical ($380$ nm $< \lambda < 930$ nm) spectrograph on the WIYN 3.5-m telescope at Kitt Peak National Observatory . NEID commissioning began following delivery to the observatory in late 2019 and concluded in May 2021, and science operations have been underway since Summer 2021. 

To maximize the scientific return of NEID, we have thought carefully about the manner in which our observations are conducted; poorly designed observing and operating strategies, at either the individual level or the facility level, can easily negate the advantages conferred by the instrument itself.
To this end, we have designed a pair of graphical user interfaces (GUIs) with which the NEID spectrograph is controlled. We refer to these herein as the NEID Spectrograph GUI and the NEID Exposure Control GUI. These GUIs were developed during the commissioning process and tested alongside the instrument, and they have been used alongside a small suite of other GUIs to conduct all science observations over the past two years.
The Spectrograph GUI is primarily used to interface with the spectrograph itself and the Exposure Control GUI allows for more detailed monitoring of the NEID exposure meter, though some controls for the spectrograph and its various clients, in addition to the NEID port adapter\cite{Logsdon2018}, appear across both GUIs. These GUIs give observers access to necessary elements of the instrument control software\cite{Bender2016} (ICS) and the nightly observing queue, streamlining the observing process. We describe the design and functionality of each GUI in greater detail in the following sections. We also describe general observing procedures for NEID and recent updates to the ICS.

\section{NEID ICS}

NEID utilizes an expanded version of the Instrument Control Software package TIMS (Twisted Instrument Control System) that was originally built for the Habitable-zone Planet Finder spectrograph\cite{Bender2016}. TIMS leverages the Twisted Python software stack to provide reliable asynchronous communication between instrument hardware systems. The system utilizes a classic server-client architecture. A server process acts as a relay to client processes, running on the main instrument computer or on other computers connected over the facility network. Connections between clients and the server use TCP/IP communication protocols. Each client connects directly to one or more hardware devices via custom drivers that we have written in python for each individual hardware device used by the spectrometer. These drivers act like individual servers and expose the native communication protocol for their hardware device. We have implemented abstracted python driver objects for a variety of communication protocols, including serial, IP multicast, TCP/IP, ZeroMQ, XML-RPC, and SMTP, which make it straightforward to implement a new driver for a new piece of hardware in a standardized format. Special methods within each driver facilitate polling of various hardware states of interest at fixed time intervals. Each driver writes the result of these queries to running telemetry logs stored on disk, and also maintains an internal state machine in memory with the information that can be queried from any client on-demand without needing to physically re-query the hardware. For example, a pressure gauge might have its pressure polled at 1 Hz by its driver, with the latest value and corresponding timestamp stored in the driver's state machine. Any client desiring the latest pressure reading only needs to query the state machine and check for staleness of the timestamp, rather than re-query the hardware itself. Some drivers are listening to protocols such as multicast or ZeroMQ for facility telemetry that is being widely broadcast over the network. They cache a local copy of this telemetry in the driver's state machine, making it available for other TIMS clients on-demand.

At the center of the NEID hardware control system is a Dell Poweredge R730 server running the CentOS 7.x operating system. This computer runs the TIMS software, the GUI interfaces, and a variety of other monitoring and data handling housekeeping systems. The computer is connected to 35 unique hardware devices that control the spectrometer, spanning the gamut from simple (e.g., pressure gauges, DACs, temperature monitors) to complex (e.g., detector controllers, including the STA Archon and the Andor Newton, and the MenloSystems Astrocomb). Many devices with low data rates are connected to the computer via RS232-over-USB converters. We utilize two flavors of these adapters: the Sabrent CB-FTDI for standard connections, and the Startech ICUSB232FTN for devices requiring a null modem connection. USB device mounting is controlled by the Udev system, which ensures that any given device always mounts to the same port, regardless of what physical port it is connected to.

We have distributed the hardware devices over 11 TIMS clients, grouped by function. Some, such as the Enviro and Calib clients, are responsible for a large number of devices (11 and 14 respectively) that are polled at low rates. Others only control a single critical device, such as the Archon client that is responsible for the main STA Archon CCD controller and connected via a dedicated GigE TCP/IP internal network. Many devices, such as temperature control, monitor and poll multiple channels, and so TIMS is typically logging to nearly 200 individual telemetry files simultaneously, some of which contain multiple records. The system has proven to be extremely robust. It's distributed and asynchronous communication means that if an individual hardware device goes offline, the remaining devices are unaffected. In addition to the regular telemetry logs, TIMS maintains a series of status and alert message states that are distributed over email and text message to selected recipients. For example, status notifications for the automated daily LN2 fills are sent to a small group of instrument team members and WIYN facility staff. Health alerts that detect a loss of facility electrical power are sent to a much wider distribution list, to facilitate a rapid response from whomever is available on-site.

\section{GUI Software Design}

The NEID Spectrograph GUI and the Exposure Control GUI are in and of themselves TIMS clients. They utilize the PyQt4 reactor, rather than the standard Twisted reactor, but otherwise connect to the TIMS infrastructure like any other client. This means that the GUIs are able to directly access the information cached in the state machines of individual NEID drivers via those drivers' respective clients. Each GUI uses a series of regular polling queries to gather the current state of hardware relevant to the observer, and the results of those queries are reflected in the GUI buttons and text boxes. The GUI can also send commands on demand to hardware, for example to move a mechanism in the calibration bench, or to start an exposure sequence. In addition to facilitating communication with other NEID clients, integration into the ICS allows for seamless communication with other facility hardware such as the NEID port adapter and the WIYN telescope itself.

While the NEID GUIs provides an interface to the TIMS system, it is not itself actually running any processes that are required for NEID operations. An example of the utility of this design is that either GUI can be shutdown and restarted in the middle of an exposure, without interrupting that exposure in any way. Once a new instance of the GUI is launched, polling of the clients is resumed and after a few seconds the new GUI instance displays the current state of the instrument and the exposure underway.

\section{The NEID Spectrograph GUI}\label{sec:gneid}

\begin{figure} [ht]
   \begin{center}
   \begin{tabular}{c} %% tabular useful for creating an array of images 
   \includegraphics[height=15cm]{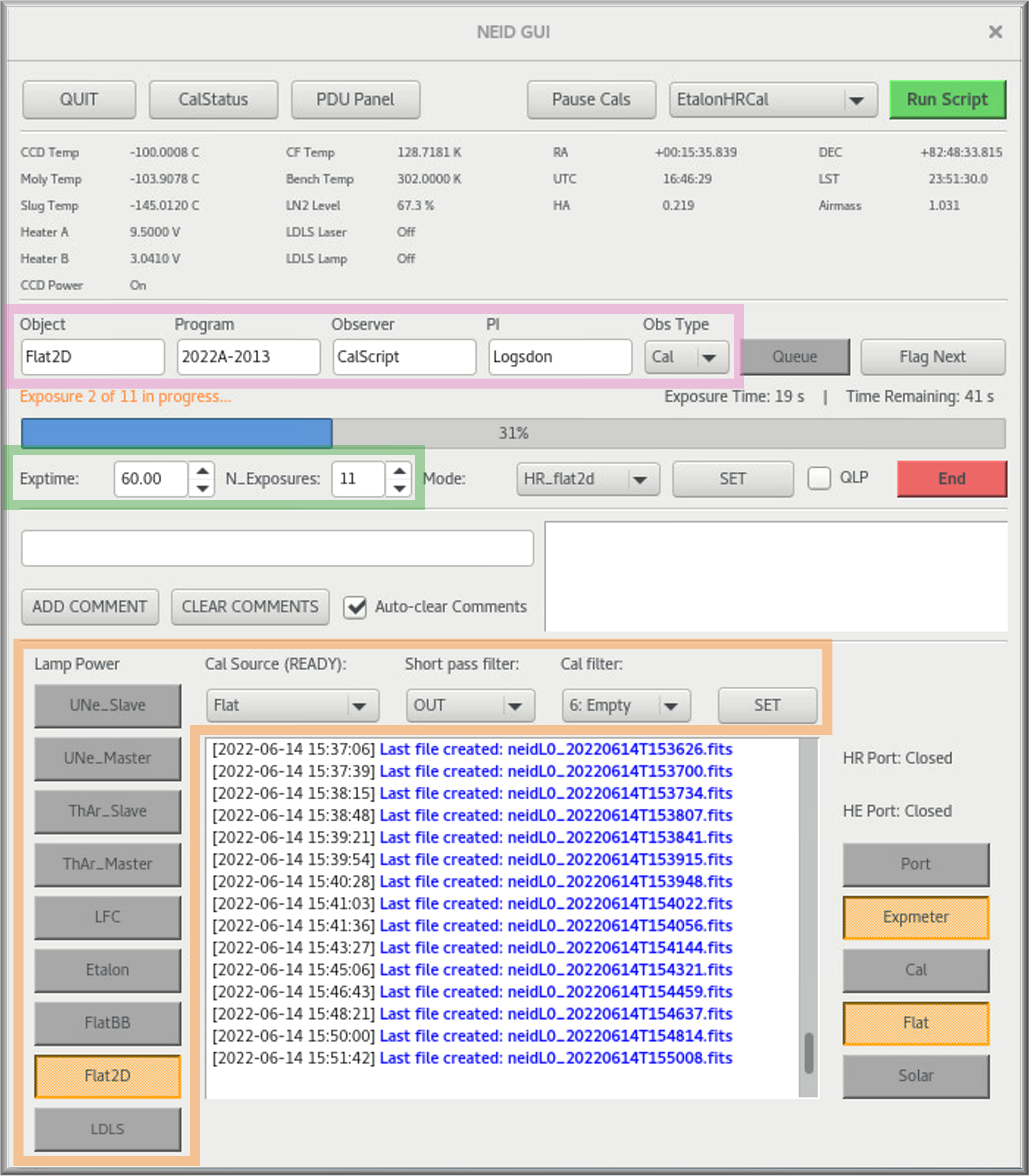}
   \end{tabular}
   \end{center}
   \caption
   { \label{fig:gneid} The NEID spectrograph GUI. This interface is used to set key observing parameters, configure the calibration bench, execute fixed exposure time observations, and monitor instrument and telescope telemetry. We indicate the header metadata section in pink, the exposure parameters in green, and the calibration bench configuration section in orange.}
\end{figure}

The NEID Spectrograph GUI (Figure \ref{fig:gneid}) serves two primary functions. First, this GUI allows observers to set the values of several metadata keywords to be included in the file headers of each observation. 
This must be done correctly to facilitate accurate ingestion and processing by the NEID Data Reduction Pipeline\footnote{https://neid.ipac.caltech.edu/docs/NEID-DRP/} (DRP), and ultimately, delivery of the final data products to the principal investigator (PI) of the appropriate observing program.

The second essential function of the NEID Spectrograph GUI is configuration of the calibration bench. Light can be sent down the NEID fibers from one of nine different light sources: a laser frequency comb, a Fabry-Perot etalon, two pairs of hollow cathode lamps (Thorium-Argon and Uranium-Neon), two flat field sources, and a laser diode light source. Users may toggle any of these sources on or off, and they can align the desired source with the beam by rotating a calibration turret (with the caveat that no more than one source can send light down the fiber at a time). In addition, a short pass cut-off filter can be flipped in or out of the light path, and a neutral density filter can be added to the path as well. The available neutral density filters are: OD 0.3, OD 0.5, OD 1.0, OD 1.3, OD 2.0.

The etalon is the only light source used during on-sky science observations; the remaining sources are used exclusively for engineering and calibration purposes.
Users may choose to include simultaneous etalon exposures when observing a star and, if so, which neutral density filter to add to the light path. Simultaneous etalon exposures are generally only recommended for observations of bright stars ($V\lesssim9$ mag), as we run the risk of introducing significant cross contamination in the science spectrum if the etalon light entering the spectrograph is much brighter than the light from the target star. When the etalon is used, an appropriate neutral density filter must therefore be selected to balance the two sources.

Although observers may configure the calibration bench and set observation parameters manually using the GUI text entry fields and dropdown menus, the GUI can also be used to configure these settings automatically, as we describe in Section \ref{sec:autoconf}.

The Spectrograph GUI also serves a few additional non-essential functions. In the top toolbar, we provide buttons that can be used to monitor the progress of calibration sequences, check the status of several instrument power distribution units (PDUs), and execute a limited number of preset scripts.
The window in the lower section of the GUI records any changes made to the spectrograph configuration and maintains a running log of recently completed exposures.
The buttons in the lower left section of the Spectrograph GUI can be used to activate or deactivate each of the calibration light sources. These buttons will turn pale orange when active and grey when inactive. We also implement a warning indicator, such that each button can turn red in the event that the associated light source is not in one of the expected, allowed states. And finally, observers may use the upper section of Spectrograph GUI to monitor -- but \textit{not} control -- other instrument vitals and confirm that they fall within acceptable ranges. Monitoring of the light sources and instrument vitals adds an appreciated level of redundancy to automated monitoring systems which trigger alerts and safe shutdown protocols, when appropriate.

\section{The NEID Exposure Control GUI}\label{sec:obsgui}

\begin{figure} [ht]
   \begin{center}
   \begin{tabular}{c} %% tabular useful for creating an array of images 
   \includegraphics[height=12cm]{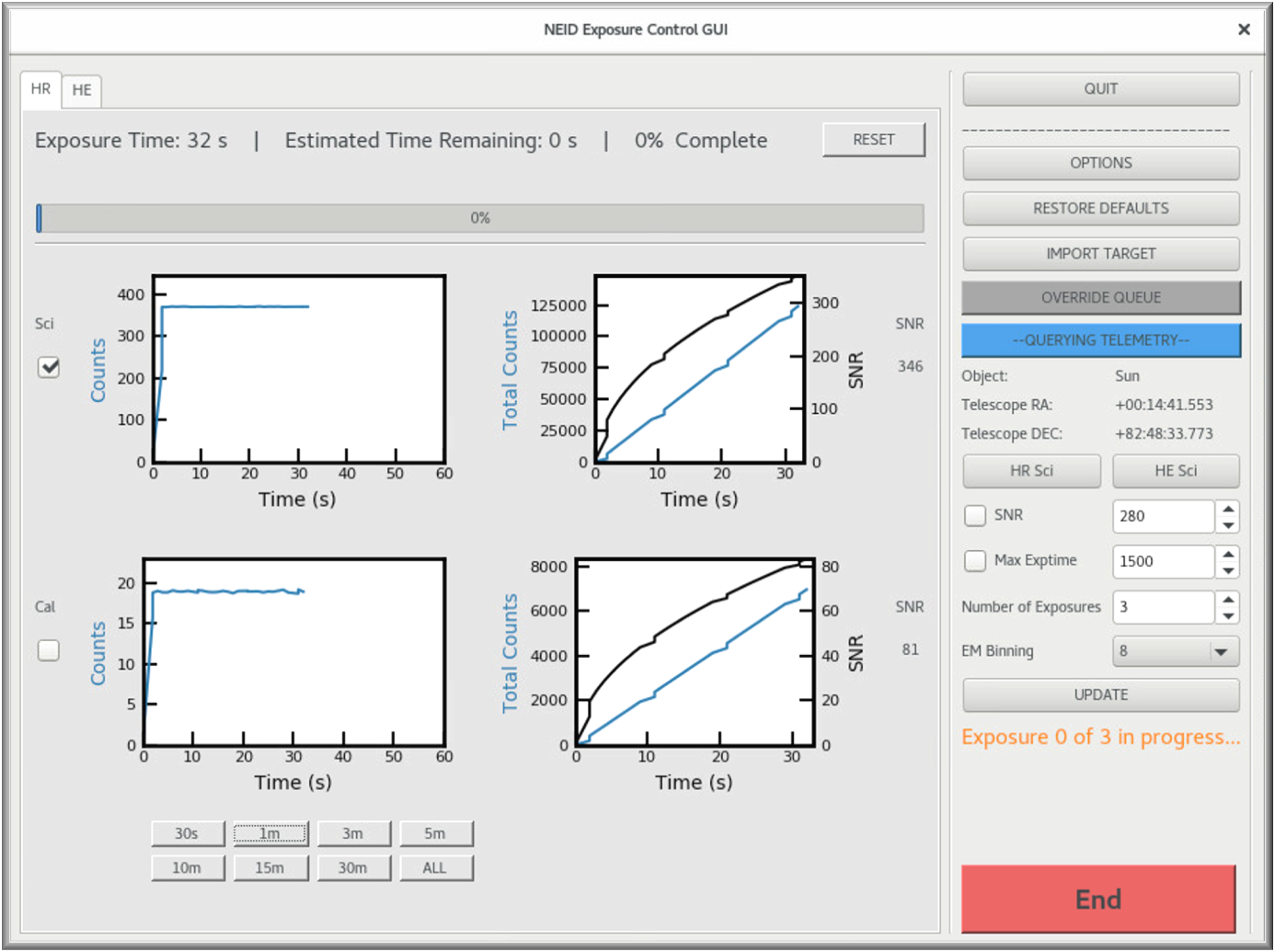}
   \end{tabular}
   \end{center}
   \caption
   { \label{fig:obsgui} NEID Exposure Control GUI. This GUI is used to execute SNR-triggered observations. Inset plots allow observers to monitor the instantaneous (left) and cumulative (right) counts obtained using the science fiber (top) and calibration fiber (bottom).}
\end{figure}

The Exposure Control GUI, shown in Figure \ref{fig:obsgui}, allows observers to monitor the progress of individual exposures and exposure sequences. The main panel of this GUI contains a set of four plots that show the instantaneous counts per second, cumulative counts, and cumulative SNR in each of the science and calibration fibers. 
We determine these values by querying the counts reported by the NEID exposure meter and applying a wavelength-dependent scaling formula to calculate the estimated counts in the spectrograph itself.
The NEID exposure meter is a detector that intercepts a tiny fraction of the light from each NEID fiber using a pickoff mirror inside the spectrograph and records the incoming flux at a 1 second cadence in up to 256 spectral bins; we refer the reader Section \ref{sec:snrtrig} for a description of how the GUI makes use of the exposure meter telemetry and to Ninan et al. (in prep.) for a more detailed description of the exposure meter itself.

As with the rest of the telemetry, the inset plots are updated at a 1 second cadence while exposures are in progress. By tracking the cumulative counts and SNR, we give the observers a sense of the present observing conditions, provided that they are familiar with the typical throughput for median conditions.
This real-time feedback also allows observers to quickly recognize and respond to guiding failures, changes in observing conditions, or other situations that would lead to a change in the instantaneous count rate. Observers can use the tabs at the top of the main panel to switch between monitoring the counts in the high resolution (HR) and high efficiency (HE) fiber sets, though only one of these sets can be used at a time.

\section{Preparing and Executing an Observation}

\subsection{Manual Configuration}\label{sec:manconf}

When setting up for a manual observation, the observer will first enter the object name, program ID, observer name, and PI name header keywords into the text fields in the Spectrograph GUI. They will then select the appropriate observation type and shutter mode from the \texttt{`Obs Type'} and \texttt{`Mode'} dropdown menus, respectively. When setting up for a calibration exposure, or when conducting a science exposure with simultaneous calibration, the observer will then configure the calibration bench. This consists of directing the turret to the correct source (with the \texttt{`Cal Source'} widget), confirming that this source is turned on (with the lamp buttons), moving the short pass filter in or out, and selecting the appropriate neutral density filter.

Finally, if necessary, relevant comments should be entered using the \texttt{`ADD COMMENT'} button; these comments will be recorded in the \texttt{`COMMENT'} field of the resulting FITS file. Comments can also be added while exposures are in progress (e.g., to add a note if the transparency changed significantly mid-exposure).

\subsection{Automatic Configuration with the Observing Queue}\label{sec:autoconf}

NEID operates on a fully queue-based observing schedule; in addition to the scheduling flexibility this provides, the queue allows us to implement features that improve operational efficiency.
With the exception of adding comments, the manual set-up procedure can be reproduced automatically for all science observations by interfacing with the NEID observing queue\footnote{https://wiyn-queuemaster.kpno.noirlab.edu/manual/index.html}. The queue contains a set of observations for which every configurable parameter, from the object name to the desired neutral density filter, has a set value.
In order to load  information into the Spectrograph GUI from the queue palette, the observer simply clicks the \texttt{`Queue'} button in the GUI.
This queue-based operating mode is preferred in most cases, as it improves both efficiency (decreasing setup time) and reliability (reducing opportunities for user error) for single observations, and it ensures consistency across epochs for observations that are repeated multiple times.

When the \texttt{`Queue'} button has been activated, the observer will be locked out from manually editing parameters using the GUI. This design choice is intended to limit opportunities for user error, as the observation parameters will have already been vetted prior to being added to the queue. But in special cases, there may be valid reasons to change certain values. To accommodate this, we include an \texttt{`OVERRIDE QUEUE'} button in the Exposure Control GUI. Selecting this button will allow observers to override queue values.

\subsection{Executing Observations with the Spectrograph GUI: Fixed exposure time}

For single exposures or sequences of exposures with fixed exposure time, the observer must set the \texttt{`Exptime'} value (in seconds) and \texttt{`N\_exposures'} value (i.e., number of exposures) in the Spectrograph GUI (Figure \ref{fig:gneid}). This can be done either manually or automatically through the queue. Once these two values are set, the observer may click the green \texttt{`Start'} button in the Spectrograph GUI to execute the observation. A sequence of \texttt{`N\_exposures'} exposures with exposure time \texttt{`Exptime'}, separated by a fixed CCD readout time of $\sim30$ seconds, will then run through to completion. The status of the current exposure is displayed in the Spectrograph GUI as a progress bar, and the status of the full exposure sequence is shown via text. While an exposure is in progress, the \texttt{`Start'} button turns into a red \texttt{`End'} button, which can be used to terminate the exposure sequence. This button is locked during CCD readout.

Fixed exposure time observations do not require the use of the Exposure Control GUI. However, observers may still find this GUI useful in monitoring the cumulative and/or instantaneous counts during each exposure as we describe in Section \ref{sec:obsgui}.

\subsection{Executing Observations with the Exposure Control GUI: SNR-triggered exposures}\label{sec:snrtrig}

Variations in the measured flux for a single target across many observations can introduce minute perturbations to the measured radial velocity as a consequence of charge transfer inefficiency\cite{Bouchy2009,Goudfrooij2006}.
For science cases that require the highest level of radial velocity precision, it is therefore important to minimize SNR variations. We accommodate this by providing the option to terminate exposures once a set SNR limit is reached rather than after a fixed duration.

Using an SNR threshold rather than a fixed exposure time also guarantees that we achieve sufficient flux in both the NEID exposure meter and the spectrograph itself while at the same time avoiding detector non-linearity effects, even on nights with unusually good or unusually poor observing conditions. For the spectrograph, this is essential when we need extremely high photon noise precision, such as when measuring sub-m s$^{-1}$ exoplanet signals or characterizing stellar variability for quiet stars. And the NEID exposure meter is designed to facilitate precise barycentric radial velocity corrections\cite{Kanodia2018} by calculating the chromatic, flux-weighted midpoint of each exposure. If the exposure meter has insufficient flux, we will be unable to compute the barycentric correction at the $<2$ cm s$^{-1}$ level specified in the instrumental error budget\cite{Halverson2016}.

\begin{figure} [ht]
   \begin{center}
   \begin{tabular}{c} %% tabular useful for creating an array of images 
   \includegraphics[height=8cm]{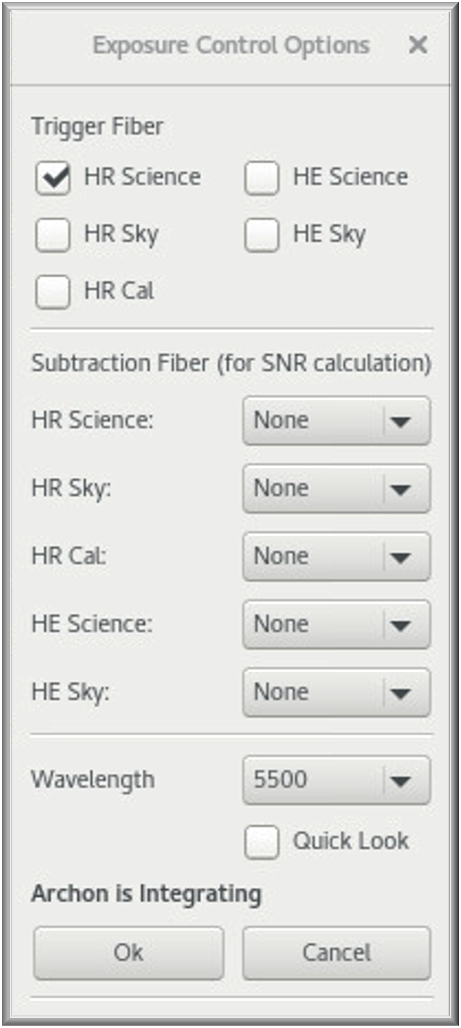}
   \end{tabular}
   \end{center}
   \caption
   { \label{fig:obs_options} Options window for the Exposure Control GUI.}
\end{figure} 

The SNR-triggered observing mode is executed through the Exposure Control GUI (Figure \ref{fig:obsgui}), though the Spectrograph GUI must still be used to configure the set up as described in Sections \ref{sec:manconf} and \ref{sec:autoconf}.  Using the Exposure Control GUI, the observer must set the SNR limit (\texttt{`SNR'}), a maximum exposure time (\texttt{`Max Exptime'}; as a safeguard against inordinately long exposures in the case of sub-optimal observing conditions), and a number of exposures \texttt{`Number of Exposures'}. We also include a separate \texttt{`OPTIONS'} menu (Figure \ref{fig:obs_options}) with which the observer can select which fiber to monitor when executing an SNR-triggered exposure and whether to subtract the flux from a second fiber (e.g., sky subtraction), as well as choose the wavelength at which the SNR will be calculated. This choice of wavelength is important when preparing observations of stars of different spectral types, for which the spectral ranges containing the bulk of the radial velocity information content differ. As with other observing parameters, the SNR, maximum exposure time, number of exposures, trigger fiber, and trigger wavelength can be configured automatically via the queue.

Once the above values have been set, the observer may execute a sequence of exposures using the green \texttt{`Start'} button at the bottom bottom of the Exposure Control GUI. 
The GUI will track the progress of the exposure and the inset plots will update to display the SNR and instantaneous and cumulative counts.
We note that these plots also update during fixed exposure time observations even though the triggers are not active.
When either the SNR threshold or the maximum exposure time is reached (whichever comes first), the exposure will be terminated, and for observations with multiple exposures, the next exposure in the sequence will be initiated following a $\sim30$ second CCD readout. We note that the SNR-triggered observing mode is only feasible for stars brighter than $V\lesssim12$ mag; for fainter stars, the low instantaneous counts measured by the exposure meter preclude reliable, real-time identification of the precise time at which a threshold is reached.

\begin{figure} [ht]
   \begin{center}
   \begin{tabular}{c} %% tabular useful for creating an array of images 
   \includegraphics[height=12cm]{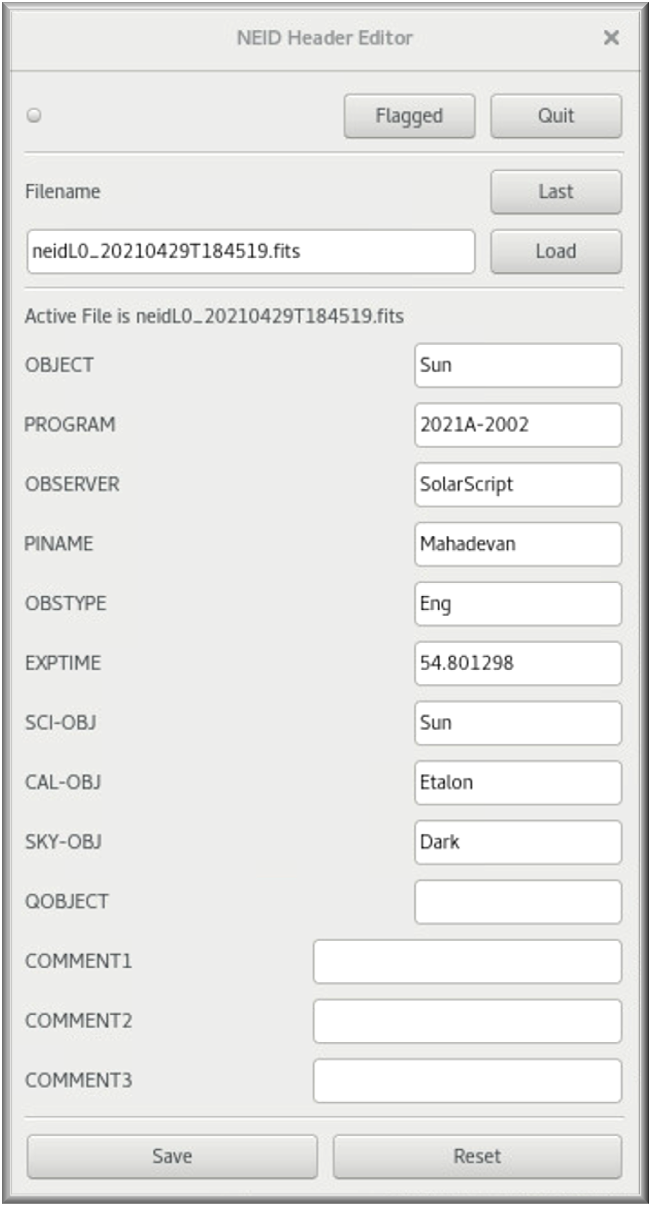}
   \end{tabular}
   \end{center}
   \caption
   { \label{fig:header_editor} NEID header editor GUI. This GUI is used to edit the headers of flagged files and mark them for re-processing.}
\end{figure} 

\section{Flagging and Correcting Data}

Erroneous header keywords can significantly complicate proper data delivery and time accounting. Should the observer note that a keyword is wrong while an exposure is in progress, they may use the \texttt{`Flag Next'} button on the right hand side of the Spectrograph GUI to flag the current observation to be corrected later on. The list of flagged files can be accessed via a separate Header Editor GUI (Figure \ref{fig:header_editor}).
After loading a flagged file, the observer may edit the text boxes for any of the header fields displayed in the Header Editor GUI to correct inaccurate information or add comments regarding the data quality, observing conditions, technical issues, or other relevant information. Fields that have been edited will be marked with an asterisk until the observer clicks the \texttt{`Save'} button, at which time the file is saved with edited header information, and the \texttt{`DATE'} keyword will be updated and the appropriate information will be sent to the server to indicate that the file will need to be reprocessed. In addition, a \texttt{`COMMENT'} field will be automatically generated specifying the fields that were changed and a \texttt{`HISTORY'} field will be added with details of how and when the changes were made.
To revert any changes that have been made without saving, the observer may use the \texttt{`Reset'} button. This will reload the header of the active file.

\section{Summary}

The NEID Spectrograph GUI and the Exposure Control GUI described here are used to carry out spectrograph operations and, in concert with several other GUIs and tools, to conduct NEID science observations.
We present a detailed breakdown of the design and functionality of these GUIs as well as a description of general observing procedures. Through interaction with the NEID observing queue, and the instrument control software, these GUIs have streamlined the observing process, reducing overheads and improving data reliability. We also touch on aspects of the design philosophy, explaining why certain features have been implemented and how they contribute to the success of the instrument. 

\acknowledgments % equivalent to \section*{ACKNOWLEDGMENTS}       
 
NEID is funded by NASA through JPL by contract 1547612. NEID is located on the WIYN 3.5-m telescope at Kitt Peak National Observatory.
The authors are honored to be permitted to conduct astronomical research on Iolkam Du’ag (Kitt Peak), a mountain with particular significance to the Tohono O’odham.

The Center for Exoplanets and Habitable Worlds and the Penn State Extraterrestrial Intelligence Center are supported by the Pennsylvania State University and the Eberly College of Science. The Pennsylvania State University campuses are located on the original homelands of the Erie, Haudenosaunee (Seneca, Cayuga, Onondaga, Oneida, Mohawk, and Tuscarora), Lenape (Delaware Nation, Delaware Tribe, Stockbridge-Munsee), Shawnee (Absentee, Eastern, and Oklahoma), Susquehannock, and Wahzhazhe (Osage) Nations.  As a land grant institution, we acknowledge and honor the traditional caretakers of these lands and strive to understand and model their responsible stewardship. We also acknowledge the longer history of these lands and our place in that history.

% References
\bibliography{references} % bibliography data in report.bib
\bibliographystyle{spiebib} % makes bibtex use spiebib.bst

\end{document}